\documentclass{sig-alternate}
\usepackage[pdftex,colorlinks=true,pdfstartview=FitV,
 linkcolor=black,citecolor=black,urlcolor=black]{hyperref}

\usepackage{xspace}
\usepackage{ifthen}
\usepackage{amssymb}
\newboolean{showcomments}
\setboolean{showcomments}{true} 
\ifthenelse{\boolean{showcomments}}
  {\newcommand{\marking}[3]{
	\fcolorbox{gray}{#3}{\bfseries\sffamily\scriptsize#1}
    {\sf\small$\blacktriangleright$\textit{#2}$\blacktriangleleft$}
   }

  }
  {\newcommand{\marking}[3]{}
   
  }

\usepackage[normalem]{ulem} 
\usepackage{xcolor}

\newcommand{\del}[1]{\textcolor{red}{\sout{#1}}} 

\newcommand{\SOCA}{Software Cartography\xspace}

\newcommand{\Codemap}{\textsc{Codemap}\xspace}

\newcommand{\eclipse}{Eclipse\xspace}
\newcommand{\ie}{\emph{i.e.},\xspace}
\newcommand{\eg}{\emph{e.g.},\xspace}
\newcommand{\etal}{\emph{et al.}\xspace}

\begin{document}

\conferenceinfo{ICSE 2010,}{Cape Town, South Africa}
\title{Towards Improving the Mental Model of Software Developers through Cartographic Visualization}
\numberofauthors{3} 
\author{
\alignauthor
Adrian Kuhn\\
       \affaddr{Software Composition Group}\\
       \affaddr{University of Bern, Switzerland}\\
       \email{http://scg.unibe.ch/akuhn}
\alignauthor
David Erni\\
       \affaddr{Software Composition Group}\\
       \affaddr{University of Bern, Switzerland}\\
       \email{http://www.deif.ch}
\alignauthor
Oscar Nierstrasz\\
       \affaddr{Software Composition Group}\\
       \affaddr{University of Bern, Switzerland}\\
       \email{http://scg.unibe.ch/oscar}
}
\date{30 July 1999}

\maketitle
\begin{abstract}
Software is intangible and knowledge about software systems is typically tacit. The mental model of software developers is thus an important factor in software engineering. 

It is our vision that developers should be able to refer to code as being ``up in the north'', ``over in the west'', or ``down-under in the south''. We want to provide developers, and everyone else involved in software development, with a \emph{shared}, \emph{spatial} and \emph{stable} mental model of their software project. We aim to reinforce this by embedding a cartographic visualization in the  IDE (Integrated Development Environment). The visualization is always visible in the bottom-left, similar to the GPS navigation device for car drivers. For each development task, related information is displayed on the map. In this paper we present \Codemap, an eclipse plug-in, and report on preliminary results from an ongoing user study with professional developers and students.
\end{abstract}

\keywords{Architecture visualization, Human Factors, Mental Model, Spatial Representation, Software Development, Software Visualization, Tool Building.}

\section{Introduction}

\begin{quote}
\emph{``If you write software knowledge down, it becomes immediately stale because it keeps changing in the developer's mind.''} -- Andrew Ko
\end{quote}

Software is intangible and knowledge about software systems and their architecture is often tacit. The mental model of software developers is thus an important factor in software engineering. In our work, we aim to provide developers with tool support to establish a better mental model of their work. We do so by embedding a cartographic visualization, called \textsc{Codemap}, in their IDE (Integrated Development Environment). We do not aim to document or visualize the present mental model of developers, rather it is our goal that developers arrive at a better mental model based on the spatial visualization provided by our tool. This is motivated by the observation that the representation of source code in the IDE often impacts the mental model of developers. Compare for example the mental model held by an \eclipse developer with that of an {\tt emacs} or {\tt vim} user, or with the even more diverging mental model of development in explorative runtime systems such as Smalltalk and Self \cite{Sand88a}.


DeLine observed that developers are consistently lost in source code and that using textual landmarks only places a large burden on cognitive memory  \cite{Deli05a}. DeLine concluded that we need new visualization techniques that allow developers to use their spatial memory while navigating source code. He proposed four desiderata that should be satisfied by spatial software navigation~\cite{Deli05b}. In our most recent work~\cite{Kuhn09x} we generalized and extended this list as follows:

\begin{enumerate}
\setlength{\itemsep}{0.5ex}%
\setlength{\parskip}{0.5ex}%
\item The visualization should show the entire program and be continuous.
\item The visualization should contain visualization landmarks that allow the developers to find parts of the system perceptually, rather than relying on naming or other cognitive feats.
\item The visualization should remain visually stable as the system evolves (both locally and across distributed version control commits). 
\item The visualization should be capable of showing global information overlays.
\item Distance in the visualization should have a intuitive though technically meaningful interpretation. 
\end{enumerate}

We implemented the \Codemap tool as a proof-of-concept prototype of \SOCA. The prototype is open-source and available as an Eclipse plug-in\footnote{\url{http://scg.unibe.ch/codemap}}. 
At the moment, we are evaluating our approach in an ongoing controlled experiment with professional developers and students. 

~

The remainder of this paper is structured as follows: 
\autoref{sec:tasks} enumerates the developments tasks that are supported by the \Codemap plug-in. 
\autoref{sec:steps} presents the codemap algorithm and its recent improvements. 
\autoref{sec:discussion} discusses preliminary results from the ongoing evaluation.
\autoref{sec:related} discusses related work.
\autoref{sec:future} concludes with remarks on future work.

\section{Software Cartography}
\label{sec:tasks}

\SOCA uses a spatial visualization of software systems to provide software development teams with a stable and shared mental model. Our cartographic visualization is most useful when it supports as many development tasks as possible. Therefore we integrated \SOCA in the IDE so that a map of the software system may always be present and may thus support as many development tasks as possible. 

At the moment, the \Codemap plug-in for \eclipse supports the following tasks:\footnote{New features are added on a weekly base, please subscribe to \url{http://twitter.com/codemap} to receive latest news.}

\begin{figure}
\begin{center}
  \includegraphics[width=\linewidth]{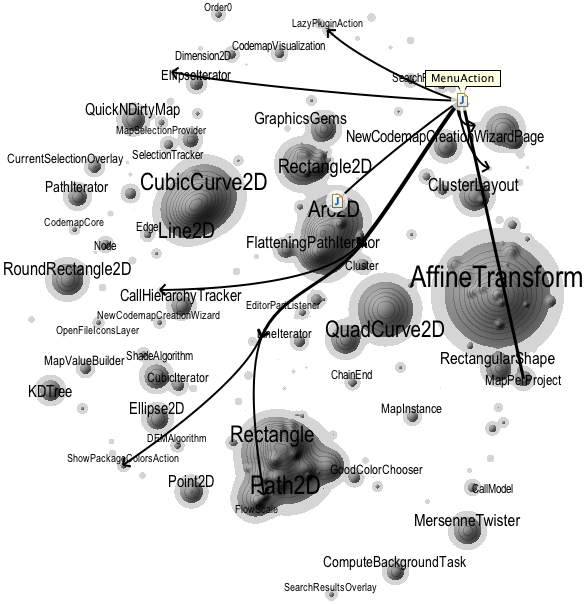}
\end{center}
    \caption{\emph{Thematic codemap of a software system, here the \Codemap tool itself is shown. Arrow edges show incoming calls to the {\tt \#getSettingOrDefault} method in the {\tt MenuAction} class, which is currently active in the editor and thus labeled with a pop-up.}}
    \label{fig:awesome}
\end{figure}

\begin{itemize}
\item Navigation within a software system, be it for development or analysis. \Codemap is integrated with the package explorer and editor of \eclipse. The selection in the package explorer and the selection on the map are linked. Open files are marked with an icon on the map. Double clicking on the map opens the closest file in the editor. When using heat map mode, recently visited classes are highlighted on the map.

\item Comparing software metrics to each other, \eg to compre bug density with code coverage. \Codemap is hooked into several \eclipse plug-ins in order to display their results on the map alongside the regular views. The map displays search results, compiler errors, and (given the \textsc{Eclemma} plug-in is installed) test coverage information. More information can be added through an extension point.

\item Social awareness of collaboration in the development team. \Codemap can connect two or more \eclipse instances to show open files of other developers. Colored icons are used to show the currently open files of all developers. Icons are colored by user and updated in real time.

\item Help to understand a software system's domain. The layout of \Codemap is based on structure \emph{and} vocabulary, since we believe that a mental model of software should transcend structural artifacts. Labels on the map are not limited to class names, but include automatically retrieved keywords and topics.

\item Exploring a system during reverse engineering. \Codemap is integrated with \eclipse's structural navigation functions such as search for callers, implementers, and references. Arrows are shown for search results. We apply the \textsc{Flow Map} algorithm \cite{Phan05a} to  avoid visual clutter by merging parallel arrow edges. \autoref{fig:awesome} shows the result of searching for calls to the {\tt \#getSettingOrDefault} method in the {\tt MenuAction} class .
\end{itemize}

\section{The Codemap Algorithm}
\label{sec:steps}

\begin{figure*}
\begin{center}
  \includegraphics[width=\linewidth]{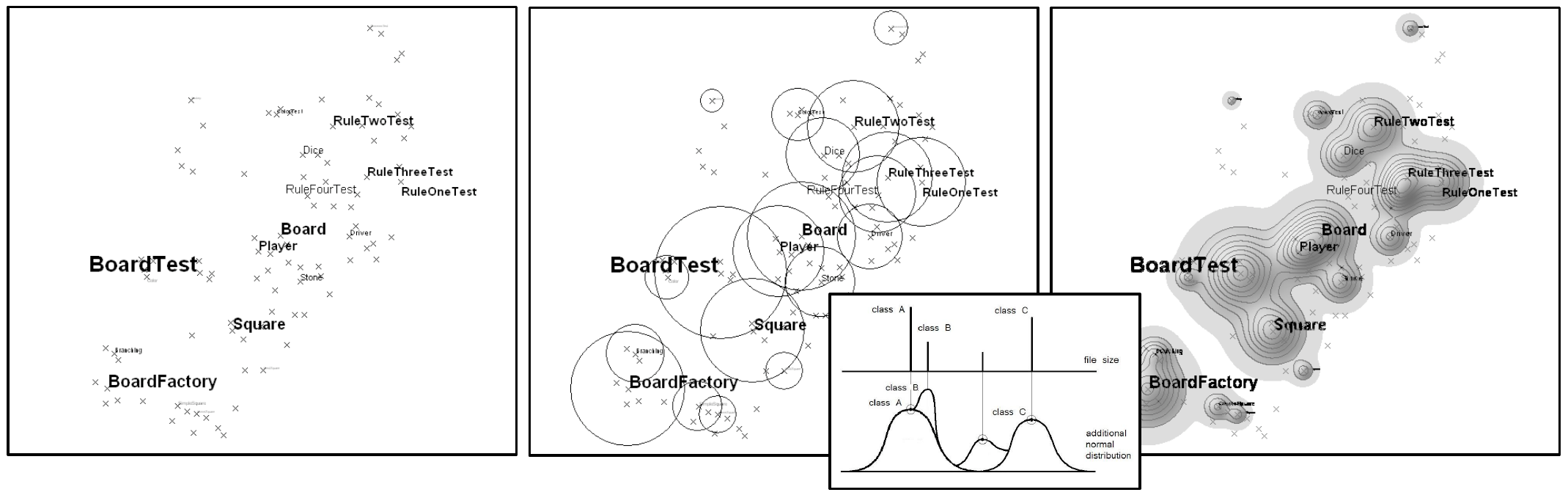}
\end{center}
    \caption{\emph{Construction steps of a software map, from left to right: 1) 2-dimensional embedding of files on the visualization pane; 2.a) circles around each file's location, based on class size in KLOC; 2.b) each file contributes a Gaussian shaped basis function to the elevation model according to its KLOC size; the contributions of all files are summed up; 3) fully rendered map with hill-shading, contour lines, and filename labels.}}
    \label{fig:steps}
\end{figure*}

\autoref{fig:steps} illustrates the construction of a software map. The sequence of the construction is basically the same as presented in previous work~\cite{Kuhn08b,Kuhn09x}. However, preliminary results from an ongoing user study have already led to a series of improvements which are discussed below.

\textbf{2-Dimensional Embedding:}
A metric distance is used to compute the pair-wise dissimilarity of software artifacts (typically source code files). A combination of the \textsc{Isomap} algorithm \cite{Tene00a} and \textsc{Multidimensional Scaling} \cite{Borg05a} is used to embed all software artifacts on the visualization pane\footnote{Different to previous work \textsc{Latent Semantic Indexing} is not applied anymore; it has been found to have little impact on the final embedding, if at all.}. The application of \textsc{Isomap} is an improvement over previous work in order to assist \textsc{MDS} with the global layout. The final embedding minimizes the error between the dissimilarity values and the visual distances.

Early prototypes of \Codemap used a distance metric that was based on lexical similarity only. However, our user study revealed that developers tend to interpret visual distance as a measure of structural dependencies, even though they were aware of the underlying lexical implementation. Based on this observation, we developed an improved distance metric that takes both lexical similarity and structural distance (based on the ``Law of Demeter'' \cite{Lieb88a}) into account. 

\textbf{Digital Elevation Model:} In the next step, a digital elevation model is created. Each software artifact contributes a Gaussian shaped basis function to the elevation model according to its KLOC size. The contributions of all software artifacts are summed up and normalized. 

Using KLOC leads to an elevation model where large classes dominate the codemap. Observations from our user study indicate that this might be misleading since developers tend to interpret size as a measure of impact. Based on this observation, we implemented a set of different impact metrics~\cite{Lanz06a} and plan to evaluate them in a fresh user study.

\textbf{Cartographic rendering:} In the final step, hill-shading is used to render the landscape of the software map. Please refer to previous work for full details~\cite{Kuhn08b}. Metrics and markers are rendered in transparent layers on top of the landscape. Users can toggle separate layer on/off and thus customize the codemap display to their needs.

\section{PRELIMINARY RESULTS}
\label{sec:discussion}

At the moment, we are evaluating our approach in an ongoing controlled experiment with professional developers and students. The scenario of the experiment is first contact with an unknown closed-source system. Participants have 90 minutes to solve 5 exploratory tasks and to fix one bug report. After the experiment, participants are asked to sketch a drawing of their mental map of the system. 

Preliminary results are mixed and challenge our assumptions on how developers would use the \Codemap tool. Most importantly it became quickly apparent that we should revise our initial assumption that lexical similarity is a valid dissimilarity metric for the cartographic layout. Sofar, all participants tend to interpret visual distance as a measure of structural dependencies---even though they were aware of the underlying lexical implementation!

Developers intuitively expect that the map meets their mental model of the system's architecture. Since this was not given, developers were not able to take advantage of the map's consistent layout. So \eg even though north/south and east/west directions had clear (semantic) interpretations, developers did not navigate along these axes. Based on this observation we started to work on \emph{anchored multidimensional scaling} such that developers may initialize the map to their mental model. Once a map has been initialized to a developer's mental model, the map is more likely to start co-evolving with and influencing the developer's mental model. Anchored MDS allows the developer to define anchors which influence the layout of the map \cite{Buja08a}. Any software artifact can be used as an anchor, even those not present on the map as for example external libraries. With this future layout algorithm, developers may \eg arrange the database layer in the south and all UI layer in the north using the respective libraries as anchors.

Another observation was that inexperienced developers (\ie students) are more likely to find the map useful than professional developers. That was not unexpected, since to power users \emph{any} new way of using the IDE is likely to slow them down, and conversely to beginners \emph{any} way of using the IDE is  novel. The only exception to this observation was \Codemap's search bar, a one-click interface to \eclipse's native search, that was used by all participants but one.

In general, participants reported that \Codemap was most useful when it displayed search results, callers, implementers, and references. A participant reported: \emph{``I found it very helpful that you get a visual clue of quantity and distribution of your search results''}. In fact, we observed that that participants almost never used the map for direct navigation but often for search and reverse engineering tasks.
\section{RELATED WORK}
\label{sec:related}

Most closely related to the work in this paper is the work on spatial representations of code\footnote{\url{http://research.microsoft.com/en-us/projects/spatialcode/}} by the HIP (Human Interfaces in Programming) group of Microsoft Research. 

DeLine proposed four desiderata that should be satisfied by spatial software navigation \cite{Deli05b}. In the same work \emph{Software terrain maps} are presented, which satisfy the properties \#1 and \#4 (\ie continuous space and global overlays). 

Venolia and Cherubini ran a series of surveys and interviews on why and how developers use visual depictions of their code, \eg \cite{Cher07b}. They found that diagrams that document design decisions were often externalized in temporary drawings and then subsequently lost. Most of the diagrams had a transient nature because of the high cost of changing whiteboard sketches to electronic renderings. 

Code Canvas by Rowan \cite{Rowa09a}, also with the HIP group, is a zoomable UML view that drops levels of details as you zoom out, and reveals code editors as you zoom in.

Using 2-dimensional embedding to visualize information based on the metaphor of cartographic maps is by no means a novel idea. \emph{Topic maps}, as they are called, have a longstanding tradition in information visualization. In fact, the work in this paper was originally inspired by media reports on Hermann and Leuthold's work on the political landscapes of Switzerland \cite{Herm03a}. ThemeScape is the best-known example of a visualization tool that uses the metaphor of cartographic maps. Topics extracted from text documents are organized into a visualization based on topical distance and surface height corresponds to topical frequency \cite{Wise95b}. 

In the software visualization literature however, topic maps are rarely used. Except for the use of graph splatting in RE Toolkit by Telea \etal \cite{Tele03a}, we are unaware of their prior application in software visualization. 

A number of software visualization tools have adopted metaphors from cartography. Typically these tools are part of a\del{n} reverse-engineering approach based on extracted models that abstract away from source code. Thus, these tools cannot be used to read source code or develop software. The two most comprehensive of these tools are:

CodeCity is an explorative environment based on the city metaphor \cite{Wett07b}. CodeCity employs the nesting level of packages for their city's elevation model, and uses a modified tree layout to position packages and classes. Order is based on size, so the layout is not stable over time. CodeCity is not integrated into an IDE, but built in top of the Moose reverse-engineering platform that offers post-mortem analysis of abstract models only.

VERSO is an explorative environment that is also based on the city metaphor \cite{Lang05a}, very similar to CodeCity. VERSO employs a treemap layout to position their elements, which provides a more stable layout. However, VERSO is also limited to post-mortem analysis of abstract models only.

\section{Remarks on Future Work}
\label{sec:future}

This paper presents \SOCA, a spatial representation of software. Our approach is supposed to help developers with a better mental model of their software systems that is stable over time and shared with team mates. Preliminary results from an ongoing user study led to the revision of our assumption that lexical similarity is sufficient to layout the cartographic map~\cite{Kuhn08b}. We refined our layout algorithm based on that conclusion. The new layout is based on both lexical similarity and the ideal structural proximity proposed by the ``Law of Demeter''.

As future work, we can identify the following promising directions:
\begin{itemize}
  \item Software maps at present are largely static.
  We envision a more interactive environment in which the user can ``zoom and pan'' through the landscape to see features in closer detail, or navigate to other views of the software.
  \item Selectively displaying features would make the environment more attractive for navigation. Instead of generating all the labels and thematic widgets up-front, users can annotate the map, adding comments and waymarks as they perform their tasks.
  \item Orientation and layout are presently consistent for a single project only.
  We would like to investigate the usefulness of conventions for establishing consistent layout and orientation (\ie ``testing'' is North-East) that will work across multiple projects, possibly within a reasonably well-defined domain.
\end{itemize}

\textbf{Acknowledgments:} We gratefully acknowledge the financial support of the Swiss National Science Foundation for the project ``Bringing Models Closer to Code'' (SNF Pro ject No. 200020-121594, Oct. 2008 -- Sept. 2010).

{\small 
\bibliographystyle{abbrv}
\bibliography{local}}

\end{document}